# Discovery of Very High Energy $\gamma$-ray emission from the extreme BL Lac object H2356-309 with H.E.S.S.

S. Pita[a], W. Benbow[b], L. Costamante[b], A. Djannati-Ataï[a],
D. Horns[c], M. Ouchrif[d], M. Tluczykont[e], for the H.E.S.S. collaboration
*(a) APC (Astroparticule et Cosmologie), CNRS, Université Paris VII, CEA, Observatoire de Paris, Paris, France*
*(b) Max-Planck-Institut fuer Kernphysik, Heidelberg, Germany*
*(c) Institute for Astronomy and Astrophysics, Tübingen, Germany*
*(d) Laboratoire de Physique Nucléaire et Physique des Particules, IN2P3/CNRS, Universités Paris VI & VII, Paris, France*
*(e) Laboratoire Leprince-Ringuet, IN2P3/CNRS, Ecole Polytechnique, Palaiseau, France*
Presenter: S. Pita (pita@in2p3.fr), fra-pita-S-abs1-og23-oral

The understanding of acceleration mechanisms in active galactic nuclei (AGN) jets and the measurement of the extragalactic-background-light (EBL) density are closely linked and require the detection of a large sample of very-high-energy (VHE) emitting extragalactic objects at varying redshifts. We report here on the discovery with the H.E.S.S. (High Energy Stereoscopic System) atmospheric-Cherenkov telescopes of the VHE $\gamma$-ray emission from H 2356−309, an extreme BL Lac object located at a redshift of 0.165. The observations of this object, which was previously proposed as a southern-hemisphere VHE candidate source, were performed between June and December 2004. The total exposure is 38.9 hours live time, after data quality selection, which yields the detection of a signal at the level of 9.0 $\sigma$ (standard deviations).

## 1. Introduction

Blazars are active galactic nuclei possessing relativistic jets pointing toward the observer. Their broad-band spectral energy distributions (SED), characterised by two large peaks, are dominated by non-thermal emission, which is believed to be produced in these jets. The low energy peak is commonly associated with synchrotron emission from relativistic electrons in local magnetic fields, while the high energy peak may have a leptonic or hadronic origin. In leptonic models, the $\gamma$-rays are produced in inverse Compton interactions between these electrons and local or external photons, whereas in hadronic models they are produced in $\pi_0$ decays resulting from the interaction of relativistic protons with the ambient radiation fields or matter.

Among blazars, BL Lacertae objects are characterised by the absence of emission-line features in the optical part of their SED. In the past, according to results from initial radio and X-ray surveys, these objects were separated into two families, the low-frequency-peaked BL Lac objects (LBL) or the high-frequency-peaked BL Lac objects (HBL), depending on whether their synchrotron emission is stronger in the radio-IR band or in the UV-X-ray band. Recently, different surveys have shown that the synchrotron peak may occur at any energy between radio and X-ray domain. The currently accepted unified picture of blazars [1] suggests that those objects with lower luminosities and higher frequency peaked synchrotron emission are efficient particle accelerators at very high energies and emit VHE $\gamma$-rays.

The detection and characterisation of the spectral properties of such objects is crucial for the understanding of the underlying acceleration mechanisms. It is also a promising tool for an indirect measurement of the diffuse EBL density, which carries unique cosmological information about galaxy formation and evolution, but whose direct measurement is subject to large uncertainties. The EBL photons in the range 0.1-20 $\mu$m have the appropriate energy to pair-produce with sub-TeV to TeV $\gamma$-rays, which are thus absorbed. Hence the signature of this absorption on the VHE spectra of blazars, which is a function of the redshift of the source, can be used as a probe of EBL density, if the intrinsic spectra are known. Thus, the understanding of the acceleration mechanisms and the measurement of the EBL density are closely linked and require the detection of a large sample of VHE emitting BL-Lac objects at varying redshifts. Until the present, only a small number of such sources have been discovered (see table in [2]). We report in this paper on the discovery, with the H.E.S.S. telescopes, of a new VHE $\gamma$-ray emitter: the extreme BL Lac object H 2356−309.

H 2356−309 was first reported as the optical counterpart [3] of the X-ray source 4U0009-33, discovered by the satellite UHURU in the early 1970's [4]. It was later associated with the X-ray source 1H2351-315 detected by the Large Area Sky Survey Experiment on board the HEAO satellite [5]. The host galaxy has been resolved as elliptical by different experiments [6][7][8][9]. The detection of stellar features by Falomo et al. [6] allowed the determination of its redshift, z=0.165. VLA and VLBA measurements have shown weak radio-emission [10][11]. The "extreme blazars" selected on the basis of their radio, optical and X-ray fluxes by Ghisellini [12], included H 2356−309 as a potential hard X-ray and TeV emitter. Subsequent observations carried out with BeppoSAX revealed a synchrotron peak at 1.8 keV [13], confirming this object as a strong candidate for TeV emission [14].

## 2. Observations with the H.E.S.S telescopes

The H.E.S.S. (High Energy Stereoscopic System) experiment is an array of four imaging atmospheric-Cherenkov telescopes located in Namibia (23.3° S, 16.5° E, 1835 m a.s.l.). It is dedicated to the search for $\gamma$-ray sources above 100 GeV. Each telescope is equipped with a 13 m diameter dish (focal length 15 m, mirror area 107 m$^2$) and a 960-pixel fast camera providing a 5° field of view (FoV). See [15] for more details about H.E.S.S.

The observations of H 2356−309 were performed between June and December 2004 with the full four-telescope array. Only the events for which at least two telescopes had triggered were selected by the hardware central trigger system [16]. The source has been tracked at zenith angles ranging from 7° to 54° with an average value of 20°. The data were taken in 28 minute runs using "Wobble" mode, where the source direction was positioned at ±0.5° relative to the centre of the FoV of view during observations. This increases the duty-cycle of the telescope by allowing on-source observations and simultaneous estimation of the background induced by charged cosmic rays using other parts of the FoV. The total exposure time is 38.9 hours after a run quality selection that removes runs for which the weather conditions are poor or the telescopes are not operating within specified requirements (trigger rate stability, hardware status).

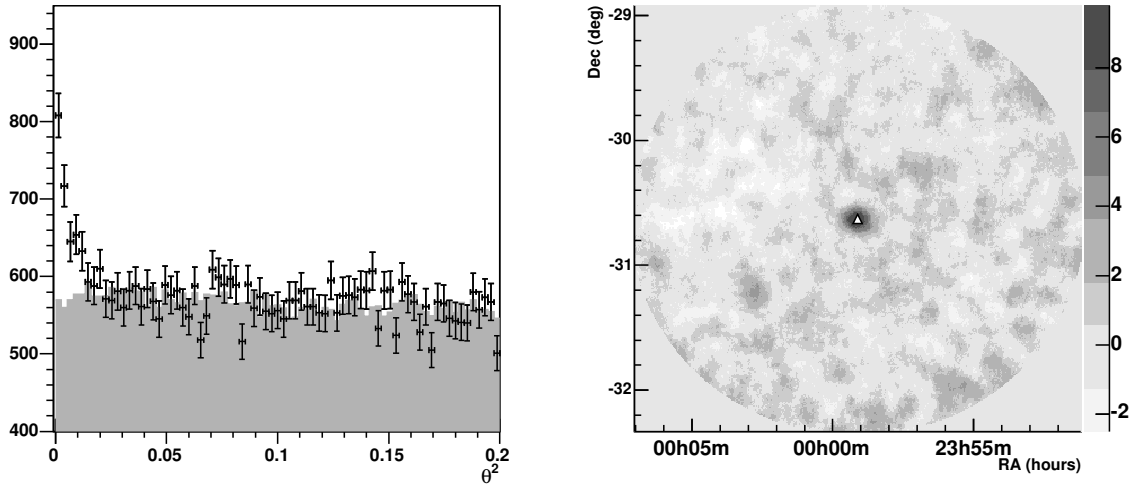

**Figure 1.** Left: The on-source (points) and normalized off-source background (grey filled) distributions of $\theta^2$ (preliminary). The excess at $\theta^2 < 0.02$ indicates VHE emission from the direction of H 2356−309. Right: A correlated two-dimensional significance map, where the H 2356−309 position is shown by a triangle. The shape of the uncorrelated excess is consistent with that of a point-like emitter located at the position of H 2356−309.

## 3. Results

The results presented below are based on the H.E.S.S. standard analysis method [15]. The left side of Figure 1 shows the on-source (points) and normalized off-source background (grey filled) distributions of $\theta^2$, where $\theta$ is the angular distance between the direction of the candidate $\gamma$-rays and the nominal position of the source. The excess at $\theta^2 < 0.02$ indicates VHE emission from the direction of H 2356−309. An excess of 645 events with $\theta^2 < 0.02\ deg^2$ is observed at a significance level of 9.0 $\sigma$ (likelihood method of Li & Ma [17]). A correlated two-dimensional significance map is shown on the right side of Figure 1. As expected, the shape of the uncorrelated excess is consistent with that of a point-like emitter located at the position of H 2356−309.

The observed integral flux above 200 GeV versus modified Julian date for each dark period is shown in Figures 2a to 2d. Figures 2a, 2b and 2c, respectively, show the night-by-night light curves for June-July, September-October and November-December 2004 periods. The month-by-month light curve is shown in Figure 2d. A fit of a constant to the data shown in Figure 2d yields a 1.4% $\chi^2$ probability, indicating flux variability on monthly time-scales. No evidence for significant variability is found on shorter time-scales. Table 1 shows the results obtained during different dark periods. The source shows a maximal activity in September, for which the excess of 210 $\gamma$-rays has a significance of 6.7 $\sigma$. The VHE spectrum of this source will be presented at the conference.

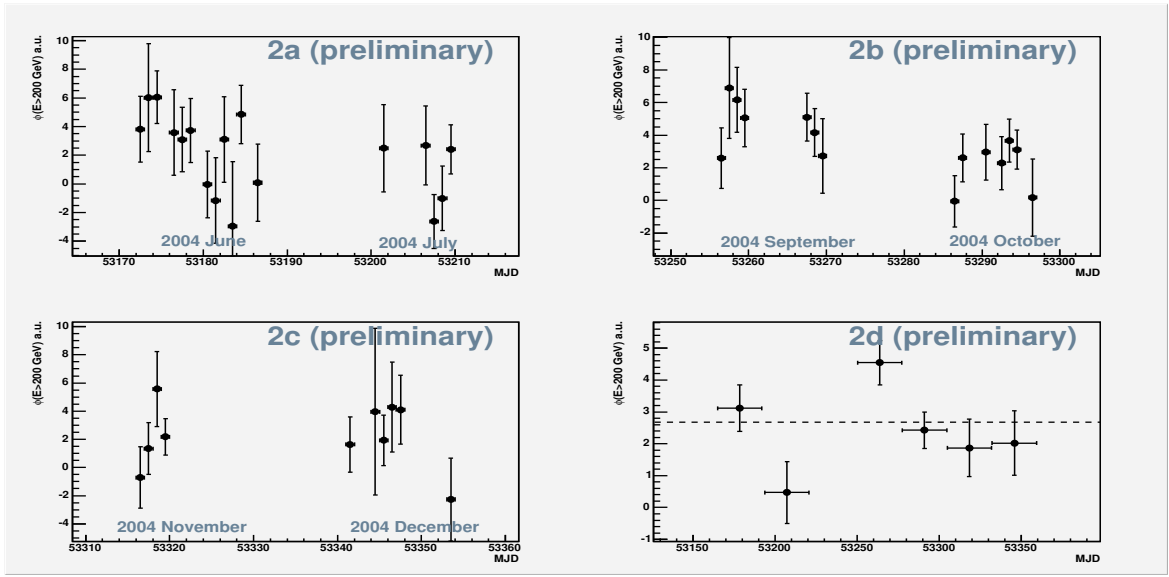

**Figure 2.** The panels show the night-by-night light for June-July (2a), September-October (2b) and November-December (2c) 2004 periods. The month-by-month light curve is given in the final panel, where the dashed line shows the result of a constant fit (preliminary).

**Table 1.** For each dark period : the number of runs, the live time, the average zenith angle, the number of *on* and *off* events passing selection cuts, the *off*-source normalisation coefficient, the excess and the corresponding significance (preliminary).

| Dark periods | Nruns | Obs. Time [$hrs$] | Zen[$deg$] | On | Off | Norm | Excess | Sig. [$\sigma$] |
|---|---|---|---|---|---|---|---|---|
| June | 21 | 7.8 | 19 | 1156 | 8032 | 0.1248 | 154 | 4.5 |
| July | 7 | 3.0 | 10 | 351 | 2737 | 0.1245 | 10 | 0.5 |
| September | 16 | 6.8 | 13 | 1001 | 6338 | 0.1248 | 210 | 6.7 |
| October | 23 | 9.8 | 15 | 1241 | 8687 | 0.1247 | 158 | 4.4 |
| November | 11 | 3.9 | 20 | 392 | 2765 | 0.1246 | 47 | 2.3 |
| December | 19 | 7.6 | 36 | 810 | 5970 | 0.1246 | 66 | 2.2 |
| Total | 97 | 38.9 | 20 | 4951 | 34529 | 0.1247 | 645 | 9.0 |

## 4. Acknowledgements

The support of the Namibian authorities and of the University of Namibia in facilitating the construction and operation of H.E.S.S. is gratefully acknowledged, as is the support by the German Ministry for Education and Research (BMBF), the Max Planck Society, the French Ministry for Research, the CNRS-IN2P3 and the Astroparticle Interdisciplinary Programme of the CNRS, the U.K. Particle Physics and Astronomy Research Council (PPARC), the IPNP of the Charles University, the South African Department of Science and Technology and National Research Foundation, and by the University of Namibia. We appreciate the excellent work of the technical support staff in Berlin, Durham, Hamburg, Heidelberg, Palaiseau, Paris, Saclay, and in Namibia in the construction and operation of the equipment.

## References

[1] Fossati, G., Maraschi, L., Celotti, A., Comastri, A., & Ghisellini, G., MNRAS, 299, 433 (1998)
[2] M. Tluczykont et al., these proceedings (2005).
[3] Schwartz et al., in LNP, edited by L. Maraschi et al. (Springer, Berling),Vol. 334, p. 211 (1989)
[4] Forman et al., ApJS, 38, 357 (1978)
[5] Wood et al., ApJS, 56 507 (1984)
[6] Falomo, AJ, 101, 821 (1991)
[7] Falomo et al., Astrophysical Journal Supp. Series, vol. 93, no. 1 (1994)
[8] Scarpa et al., Astrophysical Journal, vol. 532, Issue 2, pp. 740-815 (2000)
[9] Cheung et al., Astrophysical Journal, vol. 599, Issue 1, pp. 155-163 (2003)
[10] Laurent-Muehleisen et al., Astronomical Journal, vol. 106, no. 3, p. 875-898 (1993)
[11] Giroletti et al., Astrophysical Journal, vol. 613, Issue 2, pp. 752-769 (2004)
[12] Ghisellini, Astroparticle Physics, vol. 11, Issue 1-2, p. 11-18 (1999)
[13] Costamante et al., Astronomy and Astrophysics, v.371, p.512-526 (2001)
[14] Costamante and Ghisellini, Astronomy and Astrophysics, v.384, p.56-71 (2002)
[15] Aharonian et al. (The H.E.S.S. Collaboration), A&A, 430, 865 (2005)
[16] Funk et al., Astroparticle Physics 22, 285-296 (2004)
[17] Li, T. & Ma, Y., ApJ, 272, 317 (1983)